# The accumulation of knowledge with intra-industry knowledge spillovers: A competition game and the Nash equilibrium based on firm cost minimisation

Vasilios Kanellopoulos

University of Birmingham

School of History and Cultures

Edgbaston, Birmingham B15 2TT

United Kingdom

*Corresponding Author: Vasilios Kanellopoulos e-mail: kanellopbill@gmail.com

**Abstract:** This paper examines a competition game whose key variables are the R&D efforts (e.g. R&D expenditures) and accumulated knowledge of firms located in a specific region. The most significant element of accumulated knowledge is knowledge spillovers. These are considered intra-industry as it is assumed that the firms operate within the same industry (i.e. similar types of firms) and competitors offer similar products. The present study identifies a Nash equilibrium based on firm cost minimisation. This is derived under the assumption that the firms under examination act rationally and are primarily concerned with achieving optimal outcomes - specifically, by minimising their total costs.

**Keywords**: firms, R&D efforts, accumulated knowledge, knowledge spillovers, competition game, cost minimisation

**JEL Classification:** O3, D2, C7

# 1. Introduction

This paper examines a competition game whose key variables are the R&D efforts (e.g. R&D expenditures) and accumulated knowledge of firms located in a specific region of a country. The most significant element of this accumulated knowledge is knowledge spillovers. The present study identifies a Nash equilibrium based on firm cost minimisation. This is derived under the assumption that the firms under examination act rationally and are primarily concerned with achieving optimal outcomes - specifically, by minimising their total costs. Consequently, the present research models competition through firms' cost-minimisation decisions, incorporating their R&D efforts and accumulated knowledge which includes intra-industry knowledge spillovers.

In the literature, numerous studies focus on brand competition amongst firms for market share (Bischi et al., 2000; Monahan, 1987; Monahan and Sobel, 1994). Conversely, other studies emphasise inter-firm competition in the production of high-tech products and R&D activities, represented by R&D expenditure. This type of competition is referred to as *R&D with rivalry* (Reinganum, 1981, 1982). In this context, firms aim to obtain a greater market share and/or competitive advantage over their rivals by introducing innovations ahead of the competition or by acquiring patents (see also Bischi and Lamantia, 2004).

According to Baniak and Dubina (2012), Dasgupta and Stiglitz (1980a, 1980b) propose a general model for exploring the relationship between market structure and R&D expenditure [1]. Dasgupta and Stiglitz (1980b: 267) argue that "if firms tend to imitate each other's research strategies then much of R&D expenditure may be essentially duplicative, and consequently socially wasteful. If firms engage in excessively risky projects it may lead to too fast a rate of technical progress and high degrees of industrial concentration. This last in turn may imply large losses in production efficiency". In addition, Dasgupta and Stiglitz (1980a) underline the value of competition in R&D compared to competition in the current product market. Specifically, they find that competition in the current product market reduces innovation levels relative to monopoly, whereas competition in R&D increases innovation levels beyond the socially optimal level. In contrast, Cellini and Lambertini (2004) find that, under the open-loop equilibrium[2], a large number of firms contributes to greater product differentiation, which in turn requires higher R&D investment. This finding aligns with the Arrowian (Arrow, 1962) hypothesis that more intense competition leads to higher equilibrium and optimal levels of R&D (see also Cellini and Lambertini, 2002, 2004). Under a closed-loop equilibrium, however, a large number of firms leads to lower product differentiation, thereby requiring smaller R&D investments (Cellini and Lambertini, 2004). This result is consistent with Schumpeter's (Schumpeter, 1942) idea that the harsher the competition, the lower the optimal level of R&D (Cellini and Lambertini, 2004). Finally, Spence (1984: 101) argues that "in many markets, firms compete over time by expending resources with the purpose of reducing their costs". These expenditures primarily encompass R&D and the development of new products (Spence, 1984). In this context, the key determinants of the value of cost-reducing R&D are firm profits and revenues (see also Spence, 1984).

---

[1] Baniak and Dubina (2012) note that the classical statement of the innovation competition problem is the following: The competitive firms need to decide both on how much to spend on R&D and also on which research strategies to pursue (see also Dasgupta and Stiglitz, 1980b).

---

[2] In game theory, an open-loop model is where players cannot observe the play of their opponents, as opposed to a closed-loop model, where all past play is common knowledge.

Important aspects of firms' R&D and knowledge competition concern the fact that certain industries, such as pharmaceuticals and/or software, have high R&D costs and generate numerous inventions and patents (Bischi and Lamantia, 2004). These costs often exceed standard production costs and thus inventions are protected through patents (see Bischi and Lamantia, 2004). Another significant consideration concerns knowledge spillovers. In this context, firms seek to benefit from competitors' knowledge by exploiting its public-good characteristics, namely non-excludability and non-rivalry. Here, the concept of *absorptive capacity* suggested by Cohen and Levinthal (1989, 1990) is of great importance. According to Brinkerink (2018: 295), "Cohen and Levinthal (1989, 1990) argue that firms invest in R&D to foster the absorptive capacity needed to acquire, assimilate, and exploit knowledge inputs available outside the boundaries of the firm". In addition, the introduction of asymmetries and heterogeneities among firms with respect to spillovers helps in examining the effects of different balances between public and private knowledge (see also Bischi and Lamantia, 2004).

This paper contributes new information and insights to the literature on firm competition in respect to innovation and R&D[3]. More specifically, it is the first study to model competition through firms' cost-minimisation decisions, incorporating both R&D efforts and accumulated knowledge with intra-industry knowledge spillovers. In this framework, R&D efforts and accumulated knowledge with knowledge spillovers are the basic variables of competition. In the relevant literature, some studies model competition through profit maximisation (Bischi and Lamantia, 2004; Dasgupta and

---

[3] Indicative of this, Baniak and Dubina (2012) mention that, on the level of inter-organisational interaction, the role of innovation game theory models is to define optimal competition and cooperation strategies, in particular to determine optimal R&D expenditure.

Stiglitz, 1980a, 1980b; Spence, 1984), while other studies maximise a value function (see Reinganum, 1982).

In addition, the present research assumes that firms operate within the same sector (similar establishments) and that competitors offer similar products or services. Consequently, knowledge spillovers among competitor firms in a given region are limited within the industry; in other words, they are intra-industry spillovers (Todo et al., 2011; Yao, 2006). This assumption further allows us to assume that the firms under examination have access to the same technology and face the same given input prices. In this way, we approach the cost-minimisation problem of firms by incorporating parameters. By contrast, prior studies do not address the assumption of uniform access to technology and identical input prices. Instead, they focus on the decision of the $ith$ firm (where $i = 1,2, ... , n$) and maximise the $ith$ firm profit function (Bischi and Lamantia, 2004; Spence, 1984), thereby assuming that firms have access to different technologies and face different input prices.

Another novel aspect of this research is that it employs a challenging and demanding nonlinear total cost function for solving the firm cost minimisation problem. Most prior research relies on linear cost functions or/and simple nonlinear cost functions for this. Here, more sophisticated and challenging mathematical and algebraic methods and techniques are used to reach the optimal input (R&D efforts and accumulated knowledge) prices. The final novel contribution of this research is that it tries to economically interpret the negative value of the Nash equilibrium price of accumulated knowledge, as derived from the mathematical calculations. Given that intra-industry knowledge spillovers are the most important element of accumulated knowledge, we concentrate on the public-good nature of knowledge, which satisfies the conditions of non-excludability and non-rivalry. As the unexploited portion of the new knowledge of an examined firm spills over to other firms, we consider this part as a subsidy of this firm to rival firms. To clarify this point, we develop a model demonstrating that suppliers subsidise the cost of the input (knowledge spillovers) for buyers, mainly due to the negative value of the price of accumulated knowledge in the Nash equilibrium.

The rest of the paper is structured as follows: The second section describes the model. It first examines the accumulation of knowledge with intra-industry knowledge spillovers and the case of asymmetries (heterogeneities) among firms with respect to spillovers. It then develops the competition game and derives the Nash equilibrium based on firm cost minimisation. Finally, it provides an economic interpretation of the negative value of the Nash equilibrium price of accumulated knowledge. The last section focuses on conclusions, certain limitations of the study and avenues for further research.

## 2. The model

*2.1 The accumulation of knowledge with intra-industry knowledge spillovers: The case of asymmetries (heterogeneities) among firms with respect to spillovers*

We assume a market comprising $n$ firms, where $i = 1,2,\ldots,n$. These firms exist in the $r_{th}$ region of a country and operate within the same sector (e.g. business services)[4]. In other words, competitors offer similar products or services. We also assume that these firms produce the entire output of the region and that significant R&D (knowledge) spillovers exist among them. Consequently, knowledge spillovers are mainly intra-industry (Bernstein and Nadiri, 1989; Todo et al., 2011; Yao, 2006). First, we let $x_i(t)$ denote the R&D efforts (e.g. R&D expenditures) made by firm $i$. According to Spence (1984), the accumulated knowledge of firm $i$ in period $t$ is given by (see also Bischi and Lamantia, 2004):

$$k_i(t) = x_i(t) + \sum_{j \neq i} \theta_{ij} x_j(t), \qquad (1)$$

---

[4] For example, such region is the Silicon Valley in Northern California in the USA. Silicon Valley is a global center for high technology and innovation.

In equation (1), $\sum_{j \neq i} \theta_{ij} x_j(t)$ is the sum of knowledge spillovers from firm $j$ to firm $i$ ($\theta_{ij}$) multiplied by the R&D efforts made by firm $j$ at time $t$. As a result, $\theta_{ij} \in [0,1]$ captures knowledge spillovers from firm $j$ to firm $i$. In the case of $\theta_{ij} = 0$, there are no knowledge spillovers. In other words, no knowledge from firm $j$ is transferred to firm $i$. This arises when firm $j$ protects its intellectual property rights (e.g. patents) and avoids the spilling of knowledge among competitors. When $\theta_{ij} = 1 \; \forall \; i \neq j$, then all knowledge produced by firm $j$ is absorbed by firm $i$. When $0 < \theta_{ij} < 1$, then the spillovers are imperfect (Spence, 1984). Although Spence (1984) sets $\theta_{ij} = \theta$, implying constant knowledge spillovers (homogeneity among firms with respect to spillovers), we allow for and consider asymmetries and heterogeneities among firms with respect to spillovers as in equation (1). In this case, Bischi and Lamantia (2004: 326) argue that "some firms are more efficient than others in spilling knowledge for free from competitors, or where some firms may be more efficient than others in avoiding the diffusion of the knowledge they gained as a result of their own efforts".

*2.2 A competition game and the Nash equilibrium based on firm cost minimisation*

In this context, the profits of the $i_{th}$ firm in period $t$ are given by the following equation (see Bischi and Lamantia, 2004)[5]:

---

[5] Bischi et al. (2000) use a market share attraction model which is widely used by both advertising practitioners and model builders. This family of models is based on the assumption that the only determinant of market share is the attraction which customers feel toward each alternative. More

$$\pi_i(t) = \frac{a_i x_i(t)}{a_i x_i(t) + \sum_{j \neq i} a_j x_j(t)} - C_i(t), \tag{2}$$

where $\pi_i(t)$ are the profits of the $i_{th}$ firm[6] in period $t$ where $i = 1,2,\ldots,n$. In addition, $x_i(t)$ are the R&D efforts of $i_{th}$ firm at time $t$, $\sum_{j \neq i} a_j x_j(t)$ is the sum of R&D efforts of $j_{th}$ firm at time $t$, $a_i \geq 0$ is the coefficient of R&D efforts of the $i_{th}$ firm, $a_j \geq 0$ is the coefficient of the $j_{th}$ firm and $C_i(t)$ is the total cost of the $i_{th}$ firm at time $t$. In addition, we assume that total cost, $C_i(t)$, is a function of the R&D efforts of the $i_{th}$ firm and the accumulated knowledge of firm $i$ at $t$ period. In other words:

$$C_i(t) = C_i(x_i(t), k_i(t)), \tag{3}$$

In addition, $C_i(x_i(t), k_i(t))$ is an increasing function of $x_i(t)$ when $k_i(t)$ is fixed. On the other hand, $C_i(x_i(t), k_i(t))$ is a decreasing function of $k_i(t)$ when $x_i(t)$ is fixed. In the first case, we have:

$$\frac{\partial C_i(t)}{\partial x_i(t)} \geq 0, \tag{4}$$

In the second case, we have:

$$\frac{\partial C_i(t)}{\partial k_i(t)} \leq 0, \tag{5}$$

---

formally, the market share $s_{it}$ for brand $i$ ($i = 1,2,\ldots,n$) in period $t$ is its attraction $A_{it}$ relative to the total attraction of all brands (see also Monahan, 1987; Hilbert and Wilkinson, 1994):
$$s_{it} = \frac{A_{it}}{\sum_{j=1}^{n} A_{jt}}$$
In this vein, Bischi and Lamantia (2004) support that firm $i$ can obtain a revenue that is proportional to its own effort $x_i$ and inversely proportional to the total effects of all competing firms. Here, the profit function of firm $i$ is written as:
$$\pi_i = B \frac{a_i x_i}{\sum_{j=1}^{n} a_j x_j} - C_i(x_i)$$
where $B$ denotes the fixed sales potential of the market, the term $s_i = \frac{a_i x_i}{\sum_{j=1}^{n} a_j x_j}$ represents the market share of brand $i$, with weights $a_i > 0$ whose values depend on the units used to measure efforts, and $C_i(x_i)$ is the total cost dependent on the effort $x_i$.

[6] Another expression for the profits of the $i_{th}$ firm is mentioned by Spence (1984). He characterises them as the earnings gross of R&D expenditures for firm $i$.

The inequality (4) means that the R&D efforts of firm $i$ are an investment with an immediate negative effect as they increase its total cost (see Bischi and Lamantia, 2004). The inequality (5), on the other hand, means that the accumulated knowledge of firm $i$ decreases its total cost. This beneficial effect is mainly due to the fact that the R&D spillover effects among competitors decrease the total costs of firm $i$[7].

Then, the total cost function can be written as (see also Bischi and Lamantia, 2004):

$$C_i\big(x_i(t), k_i(t)\big) = \frac{a_i x_i(t) + \beta_i}{\gamma_i k_i(t) + \zeta_i}, \qquad (6)$$

Here, total cost is proportional to the current efforts $x_i(t)$ and inversely proportional to $k_i(t)$. In this way, the total cost function satisfies the assumptions described by equation (3) and inequalities (4) and (5). In addition, Bischi and Lamantia (2004) consider another version of total cost function, which continues to satisfy the above conditions:

$$C_i\big(x_i(t), k_i(t)\big) = \frac{x_i(t)}{1 + \gamma_i k_i(t)}, \qquad (7)$$

When equation (6) is taken into account, the profit function for the $i_{th}$ firm at time $t$ is written as:

$$\pi_i(t) = \frac{a_i x_i(t)}{a_i x_i(t) + \sum_{j \neq i} a_j x_j(t)} - \frac{a_i x_i(t) + \beta_i}{\gamma_i k_i(t) + \zeta_i}, \qquad (8)$$

When equation (7) is considered, the profit function becomes[8]:

---

[7] Given that Audretsch and Belitski (2022) argue that knowledge spillovers arise from the fact that firms in a region and industry cannot fully appropriate the new knowledge they create, and this knowledge spills over to other firms, Bernstein (1988) find that both intra-industry and inter-industry spillovers reduce the average cost of production. More specifically, inter-industry spillovers lead to a higher reduction of unit costs than intra-industry spillovers (see also Bernstein, 1988).

[8] For the remainder of the paper, we will use the total cost function described in equation (7).

$$\pi_i(t) = \frac{a_i x_i(t)}{a_i x_i(t) + \sum_{j \neq i} a_j x_j(t)} - \frac{x_i(t)}{1 + \gamma_i k_i(t)}, \tag{9}$$

where parameter $\gamma_i \geq 0$ ($i = 1,2, \ldots, n$) is interpreted as a measure of efficiency in the use of the knowledge stock and the related technology to decrease costs for further innovation (see Bischi and Lamantia, 2004).

At this stage, we assume that the examined firms ($i = 1,2, \ldots, n$) act rationally and aim to minimise their costs[9]. In addition, we assume that all examined firms ($i = 1,2, \ldots, n$) have access to the same technology and face the same given input prices (Baumol and Fischer, 1978; see also Hatch, 1979). To a great degree, this arises from the fact that firms operate within the same sector. In this case, Baumol and Fischer (1978: 441) explain that "firms all have the same total cost function $C(y)$ for the production of any vector $y$ of the outputs".

Here, we consider the following production function:

$$Q(t) = f(x(t), k(t)), \tag{10}$$

where $Q(t)$ is the output of innovation or patents, $x(t)$ are the R&D efforts and $k(t)$ is accumulated knowledge. In turn, the total cost is written as:

$$C(t) = \frac{px(t)}{1 + \gamma r k(t)}, \tag{11}$$

---

[9] According to Mohajan (2022), a firm takes a rational decision to use various raw materials for minimising its production cost. In other words, firms rationally choose the most efficient combination of inputs to produce the desired output at the lowest possible cost. This is the cost minimisation problem which, in mathematical terms, is a problem of constrained optimization. The firm wishes to minimise the cost of producing a certain level of output, but it is constrained by its technological possibilities, as summarised by the production function.

In equation (11), we have added the prices of inputs[10]. More specifically, $p > 0$ is the price rate of R&D efforts, $r > 0$ is the price rate of accumulated knowledge, the cost of R&D efforts is the price rate of R&D efforts multiplied by the amount of R&D efforts ($px(t)$) and the cost of accumulated knowledge is the price rate of accumulated knowledge multiplied by the accumulated knowledge ($rk(t)$).

Then, we mathematically express the cost minimisation problem in the following way: We want to minimise total cost subject to an output target. In other words:

$$min \frac{px(t)}{1+\gamma rk(t)}, \tag{12}$$

subject to
$$Q(t) = f(x(t), k(t)), \tag{13}$$

To solve the above problem, we introduce a Lagrangian function:

$$\Lambda(x(t), k(t), \lambda) = \frac{px(t)}{1+\gamma rk(t)} - \lambda(f(x(t), k(t)) - Q(t)), \tag{14}$$

where $\lambda$ is the Lagrange multiplier. The Lagrangian function can also be written as:

$$\Lambda(x(t), k(t), \lambda) = \frac{px(t)}{1+\gamma rk(t)} - \lambda f(x(t), k(t)) + \lambda Q(t), \tag{15}$$

Then, the first-order conditions for an interior solution are as follows:

$$\frac{\partial \Lambda}{\partial x(t)} = \frac{p*1}{1+\gamma rk(t)} - \lambda \frac{\partial f(x(t), k(t))}{\partial x(t)} = 0, \tag{16}$$

or
$$\frac{\partial \Lambda}{\partial x(t)} = \frac{p}{1+\gamma rk(t)} - \frac{\lambda \partial f(x(t), k(t))}{\partial x(t)} = 0, \tag{17}$$

---

[10] Griliches (1958) argues that research and R&D spillovers assume a price and, thus, research costs occur. More specifically, he notes that any amount of money is spent for research. He (1958: 431) continues that "the moral is that, though very difficult, some sort of cost-and-returns calculation is possible and should be made. Conceptually, the decisions made by an administrator of research funds are among the most difficult economic decisions to make and to evaluate, but basically they are not very different from any other type of entrepreneurial decision".

or
$$\frac{p}{1+\gamma r k(t)} = \frac{\lambda \partial f(x(t), k(t))}{\partial x(t)}, \tag{18}$$

or
$$p * \partial x(t) = \bigl(1 + \gamma r k(t)\bigr) * \lambda \partial f\bigl(x(t), k(t)\bigr), \tag{19}$$

or
$$p^* = \frac{(1+\gamma r k(t)) * \lambda \partial f(x(t), k(t))}{\partial x(t)}, \tag{20}$$

In addition,

$$\frac{\partial \Lambda}{\partial \lambda} = -1 * f\bigl(x(t), k(t)\bigr) + 1 * Q(t) = 0, \tag{21}$$

or
$$\frac{\partial \Lambda}{\partial \lambda} = -f\bigl(x(t), k(t)\bigr) + Q(t) = 0, \tag{22}$$

or
$$Q(t) - f\bigl(x(t), k(t)\bigr) = 0, \tag{23}$$

or
$$Q^*(t) = f\bigl(x(t), k(t)\bigr), \tag{24}$$

In turn,

$$\frac{\partial \Lambda}{\partial k(t)} = p x(t)(-1)(1+\gamma r k(t))^{-2}\gamma r - \lambda \frac{\partial f(x(t), k(t))}{\partial k(t)} = 0, \tag{25}$$

or
$$-p x(t)(1+\gamma r k(t))^{-2}\gamma r - \frac{\lambda \partial f(x(t), k(t))}{\partial k(t)} = 0, \tag{26}$$

or
$$-p x(t) \gamma r \frac{1}{(1+\gamma r k(t))^2} - \frac{\lambda \partial f(x(t), k(t))}{\partial k(t)} = 0, \tag{27}$$

or
$$\frac{-p x(t) \gamma r}{(1+\gamma r k(t))^2} - \frac{\lambda \partial f(x(t), k(t))}{\partial k(t)} = 0, \tag{28}$$

or
$$\frac{-p x(t) \gamma r}{(1+\gamma r k(t))^2} = \frac{\lambda \partial f(x(t), k(t))}{\partial k(t)}, \tag{29}$$

or
$$-p x(t) \gamma r * \partial k(t) = \lambda \partial f\bigl(x(t), k(t)\bigr) * (1+\gamma r k(t))^2, \tag{30}$$

We divide both sides of equation (30) by $\gamma r$ obtaining:

$$\frac{-px(t)\gamma r * \partial k(t)}{\gamma r} = \frac{\lambda \partial f(x(t), k(t)) * (1+\gamma r k(t))^2}{\gamma r}, \tag{31}$$

or
$$-px(t) * \partial k(t) = \frac{\lambda \partial f(x(t), k(t)) * (1+\gamma r k(t))^2}{\gamma r}, \tag{32}$$

Here, we divide both sides of equation (32) by $\lambda \partial f(x(t), k(t))$ and obtain:

$$\frac{-px(t) * \partial k(t)}{\lambda \partial f(x(t), k(t))} = \frac{(1+\gamma r k(t))^2}{\gamma r}, \tag{33}$$

or
$$\frac{-px(t) * \partial k(t)}{\lambda \partial f(x(t), k(t))} = \frac{(1 + 2\gamma r k(t) + (\gamma r k(t))^2)}{\gamma r}, \tag{34}$$

or
$$\frac{-px(t) * \partial k(t)}{\lambda \partial f(x(t), k(t))} = \frac{1}{\gamma r} + \frac{2\gamma r k(t)}{\gamma r} + \frac{(\gamma r k(t))^2}{\gamma r}, \tag{35}$$

or
$$\frac{-px(t) * \partial k(t)}{\lambda \partial f(x(t), k(t))} = \frac{1}{\gamma r} + 2k(t) + \frac{\gamma^2 r^2 (k(t))^2}{\gamma r}, \tag{36}$$

or
$$\frac{-px(t) * \partial k(t)}{\lambda \partial f(x(t), k(t))} = \frac{1}{\gamma r} + 2k(t) + \gamma r (k(t))^2, \tag{37}$$

or
$$\frac{-px(t) * \partial k(t)}{\lambda \partial f(x(t), k(t))} = \frac{1}{\gamma r} + \gamma r (k(t))^2 + 2k(t), \tag{38}$$

or
$$\frac{-px(t) * \partial k(t)}{\lambda \partial f(x(t), k(t))} - 2k(t) = \frac{1}{\gamma r} + \gamma r (k(t))^2, \tag{39}$$

or
$$\frac{-px(t) * \partial k(t)}{\lambda \partial f(x(t), k(t))} - 2k(t) = (\gamma r)^{-1} + \gamma r (k(t))^2, \tag{40}$$

or
$$\gamma r (k(t))^2 + (\gamma r)^{-1} = \frac{-px(t) * \partial k(t)}{\lambda \partial f(x(t), k(t))} - 2k(t), \tag{41}$$

or
$$\gamma r (k(t))^2 + (\gamma r)^{-1} = \frac{-px(t) * \partial k(t) - 2k(t) * \lambda \partial f(x(t), k(t))}{\lambda \partial f(x(t), k(t))}, \tag{42}$$

or
$$\gamma r(k(t))^2 = \frac{-px(t)*\partial k(t)-2k(t)*\lambda \partial f(x(t),k(t))}{\lambda \partial f(x(t),k(t))} - (\gamma r)^{-1}, \qquad (43)$$

or
$$\gamma r(k(t))^2 = \frac{-px(t)*\partial k(t)-2k(t)*\lambda \partial f(x(t),k(t))}{\lambda \partial f(x(t),k(t))} - \frac{1}{\gamma r}, \qquad (44)$$

or
$$\gamma r(k(t))^2 = \frac{\gamma r[(-px(t)*\partial k(t))-2k(t)*\lambda \partial f(x(t),k(t))]}{\gamma r \lambda \partial f(x(t),k(t))} - \frac{\lambda \partial f(x(t),k(t))}{\gamma r \lambda \partial f(x(t),k(t))}, \qquad (45)$$

or
$$\gamma r(k(t))^2 = \frac{\gamma r[(-px(t)*\partial k(t))-2k(t)*\lambda \partial f(x(t),k(t))]-\lambda \partial f(x(t),k(t))}{\gamma r \lambda \partial f(x(t),k(t))}, \qquad (46)$$

or
$$\gamma r = \frac{\frac{\gamma r[(-px(t)*\partial k(t))-2k(t)*\lambda \partial f(x(t),k(t))]-\lambda \partial f(x(t),k(t))}{\gamma r \lambda \partial f(x(t),k(t))}}{(k(t))^2}, \qquad (47)$$

or
$$\gamma r = \frac{\gamma r[(-px(t)*\partial k(t))-2k(t)*\lambda \partial f(x(t),k(t))]-\lambda \partial f(x(t),k(t))}{(k(t))^2 \gamma r \lambda \partial f(x(t),k(t))}, \qquad (48)$$

or
$$\gamma r = \frac{\gamma r[(-px(t)*\partial k(t))-2k(t)*\lambda \partial f(x(t),k(t))]}{(k(t))^2 \gamma r \lambda \partial f(x(t),k(t))} - \frac{\lambda \partial f(x(t),k(t))}{(k(t))^2 \gamma r \lambda \partial f(x(t),k(t))}, \qquad (49)$$

or
$$\gamma r = \frac{[(-px(t)*\partial k(t))-2k(t)*\lambda \partial f(x(t),k(t))]}{(k(t))^2 \lambda \partial f(x(t),k(t))} - \frac{\lambda \partial f(x(t),k(t))}{(k(t))^2 \gamma r \lambda \partial f(x(t),k(t))}, \qquad (50)$$

or
$$\gamma r + \frac{\lambda \partial f(x(t),k(t))}{(k(t))^2 \gamma r \lambda \partial f(x(t),k(t))} = \frac{[(-px(t)*\partial k(t))-2k(t)*\lambda \partial f(x(t),k(t))]}{(k(t))^2 \lambda \partial f(x(t),k(t))}, \qquad (51)$$

or
$$\gamma r + \frac{1}{(k(t))^2 \gamma r} = \frac{[(-px(t)*\partial k(t))-2k(t)*\lambda \partial f(x(t),k(t))]}{(k(t))^2 \lambda \partial f(x(t),k(t))}, \qquad (52)$$

or
$$\frac{((k(t))^2 \gamma^2 r^2)+1}{(k(t))^2 \gamma r} = \frac{[(-px(t)*\partial k(t))-2k(t)*\lambda \partial f(x(t),k(t))]}{(k(t))^2 \lambda \partial f(x(t),k(t))}, \qquad (53)$$

or
$$((k(t))^2 \gamma r) * [(-px(t)*\partial k(t)) - 2k(t)*\lambda \partial f(x(t),k(t))] =$$
$$[(k(t))^2 \lambda \partial f(x(t),k(t))] * [((k(t))^2 \gamma^2 r^2)+1], \qquad (54)$$

or
$$\gamma r * [(-px(t)*\partial k(t)) - 2k(t)*\lambda \partial f(x(t),k(t))] =$$
$$\lambda \partial f(x(t),k(t))[((k(t))^2 \gamma^2 r^2)+1], \qquad (55)$$

or
$$\gamma r * [(-px(t) * \partial k(t)) - 2k(t) * \lambda \partial f(x(t), k(t))] =$$
$$\lambda \partial f(x(t), k(t))((k(t))^2 \gamma^2 r^2) + \lambda \partial f(x(t), k(t)), \qquad (56)$$

or
$$\gamma r * [(-px(t) * \partial k(t)) - 2k(t) * \lambda \partial f(x(t), k(t))] -$$
$$\lambda \partial f(x(t), k(t))((k(t))^2 \gamma^2 r^2) = \lambda \partial f(x(t), k(t)), \qquad (57)$$

or
$$\{\gamma r(-px(t) * \partial k(t)) - \gamma r 2 k(t) * \lambda \partial f(x(t), k(t)) -$$
$$\lambda \partial f(x(t), k(t))((k(t))^2 \gamma^2 r^2)\} = \lambda \partial f(x(t), k(t)), \qquad (58)$$

or
$$\gamma r \{(-px(t) * \partial k(t)) - (2k(t) * \lambda \partial f(x(t), k(t))) -$$
$$(\lambda \partial f(x(t), k(t))((k(t))^2 \gamma r)\} = \lambda \partial f(x(t), k(t)), \qquad (59)$$

We divide both sides of equation (59) by $\gamma r$ obtaining:

$$\{(-px(t) * \partial k(t)) - (2k(t) * \lambda \partial f(x(t), k(t))) - (\lambda \partial f(x(t), k(t))((k(t))^2 \gamma r)\} =$$
$$\frac{\lambda \partial f(x(t), k(t))}{\gamma r}, \qquad (60)$$

or $[(-px(t) * \partial k(t)) - (2k(t) * \lambda \partial f(x(t), k(t)))] = \lambda \partial f(x(t), k(t))((k(t))^2 \gamma r +$
$$\frac{\lambda \partial f(x(t), k(t))}{\gamma r}, \qquad (61)$$

or $[(-px(t) * \partial k(t)) - (2k(t) * \lambda \partial f(x(t), k(t)))] = \lambda \partial f(x(t), k(t))((k(t))^2 \gamma r +$
$$(\lambda \partial f(x(t), k(t)))(\gamma r)^{-1}, \qquad (62)$$

or
$$[(-px(t) * \partial k(t)) - (2k(t) * \lambda \partial f(x(t), k(t)))] =$$
$$\gamma r [\lambda \partial f(x(t), k(t))((k(t))^2 + (\lambda \partial f(x(t), k(t)))(\gamma r)^{-2}], \qquad (63)$$

Here, we divide both sides of equation (63) by $\gamma r$ and obtain:

$$\frac{[(-px(t)*\partial k(t))-(2k(t)*\lambda\partial f(x(t),k(t)))]}{\gamma r} = [\lambda\partial f(x(t),k(t))((k(t))^2 +$$

$$(\lambda\partial f(x(t),k(t)))(\gamma r)^{-2}], \tag{64}$$

or $$\frac{[(-px(t)*\partial k(t))-(2k(t)*\lambda\partial f(x(t),k(t)))]}{\gamma r} = [\lambda\partial f(x(t),k(t))((k(t))^2 +$$

$$\lambda\partial f(x(t),k(t))(\gamma r)^{-2}], \tag{65}$$

or $$\left[(-px(t)*\partial k(t))-\left(2k(t)*\lambda\partial f(x(t),k(t))\right)\right]\frac{1}{\gamma r} =$$

$$[\lambda\partial f(x(t),k(t))((k(t))^2 + \lambda\partial f(x(t),k(t))(\gamma r)^{-2}], \tag{66}$$

Here, the term $\gamma r$ can be alternatively as follows: $\gamma r = \frac{\gamma^2 r^2}{\gamma r}$

As a result, equation (66) can be also written as:

$$\left[(-px(t)*\partial k(t))-\left(2k(t)*\lambda\partial f(x(t),k(t))\right)\right]\frac{1}{\frac{\gamma^2 r^2}{\gamma r}} =$$

$$[\lambda\partial f(x(t),k(t))((k(t))^2 + \lambda\partial f(x(t),k(t))(\gamma r)^{-2}], \tag{67}$$

or $$\left[(-px(t)*\partial k(t))-\left(2k(t)*\lambda\partial f(x(t),k(t))\right)\right]\frac{\gamma r}{\gamma^2 r^2} =$$

$$[\lambda\partial f(x(t),k(t))((k(t))^2 + \lambda\partial f(x(t),k(t))(\gamma r)^{-2}], \tag{68}$$

or $$\left[(-px(t)*\partial k(t))-\left(2k(t)*\lambda\partial f(x(t),k(t))\right)\right]\gamma r\frac{1}{\gamma^2 r^2} =$$

$$[\lambda\partial f(x(t),k(t))((k(t))^2 + \lambda\partial f(x(t),k(t))(\gamma r)^{-2}], \tag{69}$$

or $$\left[(-px(t)*\partial k(t))-\left(2k(t)*\lambda\partial f(x(t),k(t))\right)\right]\gamma r\frac{1}{(\gamma r)^2} =$$

$$[\lambda\partial f(x(t),k(t))((k(t))^2 + \lambda\partial f(x(t),k(t))(\gamma r)^{-2}], \tag{70}$$

or $$\left[\bigl(-px(t)*\partial k(t)\bigr)-\bigl(2k(t)*\lambda\partial f(x(t),k(t))\bigr)\right]\gamma r(\gamma r)^{-2}=$$
$$[\lambda\partial f(x(t),k(t))((k(t))^2+\lambda\partial f(x(t),k(t))(\gamma r)^{-2}], \tag{71}$$

or $$\frac{\gamma r}{1}(\gamma r)^{-2}\left[\bigl(-px(t)*\partial k(t)\bigr)-\bigl(2k(t)*\lambda\partial f(x(t),k(t))\bigr)\right]=$$
$$[\lambda\partial f(x(t),k(t))((k(t))^2+\lambda\partial f(x(t),k(t))(\gamma r)^{-2}], \tag{72}$$

or $$\frac{\gamma r}{1}(\gamma r)^{-2}\left[\bigl(-px(t)*\partial k(t)\bigr)-\bigl(2k(t)*\lambda\partial f(x(t),k(t))\bigr)\right]=$$
$$\lambda\partial f(x(t),k(t))((k(t))^2+\lambda\partial f(x(t),k(t))(\gamma r)^{-2}, \tag{73}$$

or $$\frac{\gamma r}{1}(\gamma r)^{-2}\left[\bigl(-px(t)*\partial k(t)\bigr)-\bigl(2k(t)*\lambda\partial f(x(t),k(t))\bigr)\right]-$$
$$\lambda\partial f(x(t),k(t))(\gamma r)^{-2}=\lambda\partial f(x(t),k(t))((k(t))^2, \tag{74}$$

or $$\frac{\gamma r}{1}(\gamma r)^{-2}\{\bigl(-px(t)*\partial k(t)\bigr)-\bigl(2k(t)*\lambda\partial f(x(t),k(t))\bigr)-$$
$$\lambda\partial f(x(t),k(t))\}=\lambda\partial f(x(t),k(t))((k(t))^2, \tag{75}$$

or $$\frac{\gamma r}{1}(\gamma r)^{-2}=\frac{\lambda\partial f(x(t),k(t))((k(t))^2}{\{(-px(t)*\partial k(t))-\bigl(2k(t)*\lambda\partial f(x(t),k(t))\bigr)-\lambda\partial f(x(t),k(t))\}}, \tag{76}$$

or $$(\gamma r)^{-1}=\frac{\lambda\partial f(x(t),k(t))((k(t))^2}{\{(-px(t)*\partial k(t))-\bigl(2k(t)*\lambda\partial f(x(t),k(t))\bigr)-\lambda\partial f(x(t),k(t))\}}, \tag{77}$$

or $$\frac{1}{\gamma r}=\frac{\lambda\partial f(x(t),k(t))((k(t))^2}{\{(-px(t)*\partial k(t))-\bigl(2k(t)*\lambda\partial f(x(t),k(t))\bigr)-\lambda\partial f(x(t),k(t))\}}, \tag{78}$$

or $$\gamma r*\lambda\partial f(x(t),k(t))((k(t))^2=\{\bigl(-px(t)*\partial k(t)\bigr)-\bigl(2k(t)*\lambda\partial f(x(t),k(t))\bigr)-\lambda\partial f(x(t),k(t))\}, \tag{79}$$

or $$\gamma r=\frac{\{(-px(t)*\partial k(t))-\bigl(2k(t)*\lambda\partial f(x(t),k(t))\bigr)-\lambda\partial f(x(t),k(t))\}}{\lambda\partial f(x(t),k(t))((k(t))^2}, \tag{80}$$

or
$$r = \frac{\frac{\{(-px(t)*\partial k(t))-(2k(t)*\lambda\partial f(x(t),k(t)))-\lambda\partial f(x(t),k(t))\}}{\lambda\partial f(x(t),k(t))((k(t))^2}}{\gamma}, \qquad (81)$$

or
$$r^* = \frac{\{(-px(t)*\partial k(t))-(2k(t)*\lambda\partial f(x(t),k(t)))-\lambda\partial f(x(t),k(t))\}}{\gamma\lambda\partial f(x(t),k(t))((k(t))^2}, \qquad (82)$$

As a result, $r^*$, $p^*$ and $Q^*(t)$ are the optimum values. These values constitute the Nash equilibrium and are derived from firm cost minimisation. They also incorporate the assumption that the examined firms have access to the same technology and face the same given input prices.

*2.3 The economic interpretation of the negative value of the Nash equilibrium price of accumulated knowledge: Developing a subsidisation model for the cost of the knowledge spillovers*

The findings presented in the previous subsection focus, among else, on the negative value of the Nash equilibrium price of accumulated knowledge as described in equation (82)[11]. This result is not unexpected. First, knowledge satisfies the conditions of non-excludability and non-rivalry, which are characteristics of a public good (see Pereira de Almeida et al., 2021)[12]. Knowledge spillovers, where knowledge from one entity unintentionally benefits others, are often characterised by non-excludability, meaning it is difficult or impossible to prevent others from benefiting from this. In turn, non-rivalry means that one entity's use of the knowledge does not diminish its availability for others. The negative value of the price of accumulated knowledge in the Nash equilibrium described in equation (82) could be interpreted as a subsidy associated with a public good, with knowledge serving as that good[13].

---

[11] Equation (82) can be alternatively written as: $r^* = \frac{-\{(px(t)*\partial k(t))+(2k(t)*\lambda\partial f(x(t),k(t)))+\lambda\partial f(x(t),k(t))\}}{\gamma\lambda\partial f(x(t),k(t))((k(t))^2}$

[12] Pereira de Almeida et al. (2021) underline that knowledge has also certain characteristics of a private good.

In addition, the most significant element of accumulated knowledge is knowledge spillovers. As Zuo and Lin (2022: 2) support "the benefits of a firm's R&D can be spilled out over to competitors fairly easily (Haskel and Westlake, 2018); as a result the investing firm cannot fully capture the potential return on its R&D investment". Therefore, the unexploited part of the new knowledge spills over to other firms (see also Audretsch and Belitski, 2022). This unexploited part of new knowledge could be considered as a subsidy by a firm to other firms[14] given that this firm does not set a price for the product it supplies nor place any financial demand on other firms. This could explain the negative value of the price of accumulated knowledge in the Nash equilibrium.

To clarify this point, we divide the sample of $n$ firms. Specifically, we consider that $\frac{n}{2}$ of these firms are suppliers of the input -in our case, knowledge spillovers- while the remaining $\frac{n}{2}$ are buyers of the input (knowledge spillovers). In addition, we assume that there are not property and intellectual rights (e.g. patents) in this market and that knowledge spillovers occur freely. Undoubtedly, knowledge satisfies the conditions of both the non-excludability and non-rivalry, as previously discussed. The $\frac{n}{2}$ suppliers supply an unbounded quantity of knowledge spillovers which is fully absorbed by the $\frac{n}{2}$ buyers. Consequently, the buyers acquire the entire quantity of spillovers supplied. For analytical purposes, we assume that the quantity of knowledge spillovers supplied tends towards infinity. Under this assumption, the inverse supply function of the input is expressed as follows[15]:

$$P = 9 + \frac{a}{Q_s} = 9 + \frac{a}{\infty} = 9 + 0 = 9, \tag{83}$$

---

[13] It has been argued that R&D subsidies are an input-driven policy for supporting firms' innovation capacity, reducing the costs of R&D, improving market success rate, and increasing margin (see Zuo and Lin, 2022).

[14] In vertical related markets, there are some studies where a firm uses negative input prices to subsidy a product for other firms (see Alipranti et al. 2014; Masak and Mukherjee, 2017). More specifically, the monopoly input supplier (upstream firm) subsidises the quantity setting downstream firms' production via negative input prices. Then, the assumption of negative input prices encourages the downstream firms to buy an unbounded amount of inputs knowing that the upstream firm pays the downstream firms for each unit of input they purchase.

$\forall\, a \in \mathbb{R}$ and given that $Q_s \to \infty$ and $Q_s = Q_D$ where $Q_s$ is the quantity supplied and $Q_D$ is the demand supplied.

The price described in equation (83) is the price at which suppliers supply the input. This price simultaneously constitutes a subsidy from the $\frac{n}{2}$ suppliers to the $\frac{n}{2}$ buyers. In other words, the suppliers pay for each unit of the input purchased by the buyers, who in turn accept the above price as a subsidy from the suppliers. Given that the input in question (knowledge spillovers) exhibits public-good characteristics, the buyers do not incur the cost of the input. Instead, they benefit from the subsidy associated with the input's public nature, as provided by the suppliers.

Accordingly, the profit function for the $i_{th}$ firm can be expressed as[16]:

$$\pi_i(t) = \frac{a_i x_i(t)}{a_i x_i(t) + \sum_{j \neq i} a_j x_j(t)} - \frac{p_i x_i(t)}{\gamma_i r_i k_i(t)}, \tag{84}$$

or
$$\pi_i(t) = \frac{a_i x_i(t)}{a_i x_i(t) + \sum_{j \neq i} a_j x_j(t)} + \frac{p_i x_i(t)}{\gamma_i r_i k_i(t)}, \tag{85}$$

---

[15] Here, we can imagine the case of a perfectly elastic supply (a horizontal line on a graph) where infinite quantities of input are supplied by suppliers at the same price. These infinite quantities are also demanded at that price by the buyers.

[16] Here, we use a simpler form of the total cost function by removing the unit in the denominator in equation (85). In this case, we have also found a negative value for the price of the input (accumulated knowledge) in the Nash equilibrium, following the appropriate mathematical calculations.

given $r_i^* < 0$ (negative value of the price of accumulated knowledge in the Nash equilibrium).

Equation (85) shows that, given the public-good characteristics of the input (knowledge spillovers), buyers do not pay the cost of the input. Instead, the suppliers subsidise this cost on behalf of the buyers. This interpretation also explains the positive sign of the total cost function within the general profit function in equation (85), which arises from the negative value of the price of accumulated knowledge in the Nash equilibrium. By contrast, the sign of the total cost function within the profit function should be negative (as is standard when we examine the total cost function) reflecting that buyers and demanders should pay the cost of the input[17].

## 3. Conclusions

This paper examines a competition game whose main variables are firms' R&D efforts, represented by R&D expenditures and accumulated knowledge. The most significant element of this accumulated knowledge is knowledge spillovers. These are assumed to be intra-industry, since firms operate within the same industry and competitors offer similar products. Assuming that these firms act rationally and aim at optimal outcomes, we proceed to the cost minimisation problem and its solution, from which the Nash equilibrium of the game emerges.

The current research is novel and contributes new knowledge to the existing literature, which primarily focuses on defining optimal competition and cooperation strategies among firms in respect to innovation and R&D and, thus on determining optimal R&D efforts and expenditures. More specifically, it models competition through the cost minimisation decisions of firms and, to our knowledge, is the first to do so. While some studies model competition through firms' profit maximising effort decisions (the profit maximisation problem) and others by maximising a value function, the present study assumes that the examined firms have access to the same technology and face the same given input prices. We advance to the solution of the firm cost minimisation problem on this basis. By contrast, most studies consider that firms have access to different technologies and face different input prices, thereby analysing the decisions of the $ith$ firm and focusing on the maximisation of its profits.

---

[17] This is the case and happens when the inputs are private.

In addition, this study employs demanding nonlinear total cost functions for solving the cost-minimisation problem and applies sophisticated and challenging mathematical (e.g. Lagrangian function) and algebraic methods and techniques to determine the optimal solution and the Nash equilibrium. Finally, we develop a subsidisation model for the cost of knowledge spillovers, demonstrating that suppliers subsidise the cost of the input (knowledge spillovers) for buyers.

One limitation of the current research could be the (extreme) assumption that the quantity of knowledge spillovers supplied (offered by suppliers to the buyers as a subsidy) tends to infinity and, thus, the existence of a perfectly elastic supply. However, at the same time, it is reasonable to acknowledge that large quantities of knowledge are transferred globally on a daily basis. As highlighted by Acs et al. (2013), codified knowledge, such as patents, academic articles, and books (among others), is transmitted over long distances with rather low or negligible cost. This is a kind of subsidy of firms to other firms given that the cost paid is low or negligible. In addition, the cost of tacit knowledge (non-codified knowledge) is low, as it is transferred through personal contacts and face-to-face communication.

Another limitation of the present research is the lack of empirical work and practice, which could lead to safer conclusions about the nature of inter-firm competition regarding innovation and R&D. Here, the existence of data for firm R&D expenditures in different industries and the cost of R&D employees, along with the appropriate econometric methods could provide us with more accurate information regarding the R&D efforts and knowledge spillovers given that the total cost depends on these variables. On the other hand, the combination of public R&D (higher education and government R&D) and private R&D could also capture asymmetries and heterogeneities between firms and the public sector with respect to spillovers. The current research does not include this parameter in the suggested model.

Future research could continue to explore the mathematical modelling of inter-firm competition regarding R&D. Here, scholars could place emphasis on the profit maximisation of firms and the optimal R&D employees (another aspect of R&D besides R&D expenditures) could be derived through the profit maximisation process. In this process, future research could also employ firm turnovers from product innovation (see Kanellopoulos and Tsekouras, 2023) instead of market share attraction model in order to capture the firm's revenue. Another point that could draw the attention of future research (especially when interest turns to firms' competition regarding innovation and R&D) is the use of innovation efficiency instead of R&D efforts denoted by R&D expenditures or/and R&D employees. In this way, the innovation process and innovation game theory models take into account the efficient usage of knowledge resources and their transformation into desirable innovation output (Cruz-Cazares et al., 2013; Xu et al., 2025). To summarise the potential of the future research, scholars might also investigate the development of models explaining the potential of R&D subsidies, both from government to firms and between firms themselves. Such an approach would reaffirm that knowledge and R&D possess public-good characteristics - an aspect that should be accounted for in all future analyses.

Bernstein J. and I. Nadiri (1989), 'Research and Development and intra industry spillovers: An empirical application of dynamic duality', *The Review of Economic Studies*, 56(2), 249-267.

Bischi J. I. (2000), 'Analysis of global bifurcations in a market share attraction model', *Journal of Economic Dynamics and Control*, 24(5-7), 855-879.

Bischi J. I. and F. Lamantia (2004), 'A competition game with knowledge accumulation and spillovers', *International Game Theory Review*, 6(3), 323-341.

Brinkerink J. (2018), 'Broad search, deep search, and the absorptive capacity performance of family and nonfamily firm R&D', *Family Business Review*, 31(3), 295-317.

Cellini R. and L. Lambertini (2002), 'A differential game approach to investment in product differentiation', *Journal of Economic Dynamics and Control*, 27(1), 51-62.

Cellini R. and L. Lambertini (2004), 'Private and social incentives towards investment in product differentiation', *International Game Theory Review*, 6(4), 493-508.

Cohen W. and D. Levinthal (1989), 'Innovation and learning: The two faces of R&D', *The Economic Journal*, 99(397), 569-596.

Cohen W. and D. Levinthal (1990), 'Absorptive capacity: A new perspective on learning and innovation', *Administrative Science Quarterly*, 35(1), 128-152.

Cruz-Cazares C., C. Bayona-Saez and T. Garcia-Marco (2013), 'You can't manage right what you can't measure well: technological innovation efficiency', *Research Policy*, 42(6-7), 1239-1250.
25